# Enabling Programmable Mechanical Functions Using Mechanical Property and Geometry Tuning Approaches

Fan Liu, Xihang Jiang, Zian Jia, and Lifeng Wang*

Department of Mechanical Engineering, Stony Brook University, Stony Brook, NY 11794, USA

* Corresponding authors. E-mail address lifeng.wang@stonybrook.edu

**Abstract:**

The rapidly developing robotics industry demands structures with novel mechanical functions. Traditional approaches, developing new materials and designing new structures, face two challenges. Highly complex force-displacement functions can hardly be realized and one fabricated structure only has one or limited function. We perform theoretical calculations and FE simulations to demonstrate that the two challenges can be solved by making the mechanical property or geometry of the structure tunable. The mechanical property/geometry tuning approaches are further validated by experimental tests and the results show that multiple different and highly complex objective functions can be achieved accurately. We expect that the proposed tuning approaches will facilitate the development of advanced forms of mechanical metamaterials and a new generation of robotic hands.

## 1. Introduction

A fundamental problem in the study of solid mechanics is to determine the relationship between applied external force and the deformation of the structure. In general, the relationship between the force and the deformation can be written as $F = f(M, S, \Delta)$, where the two main factors that affect the relationship are the material property ($M$) and geometry ($S$) of the structure. By understanding this relationship, multiple structures using various materials can be designed to meet different requirements as shown in Fig.1(a). For a structure made of elastic material under small linear deformation, the force-displacement relationship is linear. This linear relationship is the fundamental theory for designing many engineering structures such as bridges and towers. But apparently, not all design requirements can be satisfied by this simple linear relationship. For instance, crash attenuators are designed to absorb the colliding vehicle's kinetic energy, therefore plastic materials are used to meet this requirement. Viscoelastic materials are used in spinal dynamic stabilization devices that require energy dissipation without permanent plastic deformation[1]. Here, by choosing different materials, different force-displacement relationships are obtained and therefore, different functions are realized.

Other than choosing different materials, another method of obtaining the target force-displacement function is designing special geometries, or in other words, programmable mechanical metamaterials[2-7]. The basic idea of programmable mechanical metamaterial is designing specific geometries to achieve novel deformation modes and programmable responses as shown in Fig.1(a). In general, there are two types of programmable mechanical metamaterials: pre-programmed metamaterials[8-12] and re-programmable metamaterials[13-17]. For pre-programmed metamaterials, the programmability of the metamaterial exits only in the design phase. Different structures with different geometries are designed to achieve different responses. And then, different functionalities are programmed into geometries of different metamaterials and cannot be altered after fabrication. For re-programmable metamaterials, by applying different constrains, different functionalities can be tuned post-fabrication. The major difference between these two types of metamaterials is that one pre-programmed metamaterial has one force-displacement function while one re-programmable metamaterial has multiple force-displacement functions.

The robotics industry is growing rapidly and has a huge demand for structures that have novel functions[18-22]. For instance, as shown in Fig.1(b), the force increases with displacement to a specific value and remains constant within a given displacement range. This force-displacement function can be used in robot hands for picking fragile items. Moreover, the blue curve in Fig.1(b) shows a force-displacement curve that increases with fluctuations, which can be used in physical therapy robots. As discussed above, by developing new materials or designing new geometric structures for programmable metamaterials, numerous novel force-displacement functions can be achieved. However, two major challenges remain unsolved: complexity and universality. Specifically, the objective force-displacement function can be highly complex, as the blue curve in Fig.2(b). Developing a new material to achieve such a complex function is hardly possible. In the meantime, designing a metamaterial with new geometric structures to achieve the complex function could be challenging and time-consuming. Another challenge is using one material/structure to achieve multiple functions. For one material or one pre-programmed metamaterial, only one force-displacement function can be achieved. Even for a re-programmable metamaterial, its force-displacement functions are limited. It is hardly possible to design a re-programmable metamaterial that has the red and the blue force-displacement curves at the same time.

In this paper, we will focus on developing two approaches that make a structure obtain multiple target force-displacement functions. We will first discuss the idea of turning the material properties with respect to the displacement and therefore obtaining the target force-displacement function. Then we will perform an experiment, in which, the target force-displacement function will be obtained by changing the temperature of the material during the loading process. After that, we will design a simple structure with a cantilever beam with moving support. We will discuss the second idea of programming the position of the moving support to obtain complex target force-displacement functions accurately. Later, we will perform FEM simulations and experimental tests to validate this approach. Finally, we will design a programmable metamaterial using the simple structure as a unit cell and show its advantage in complexity and universality.

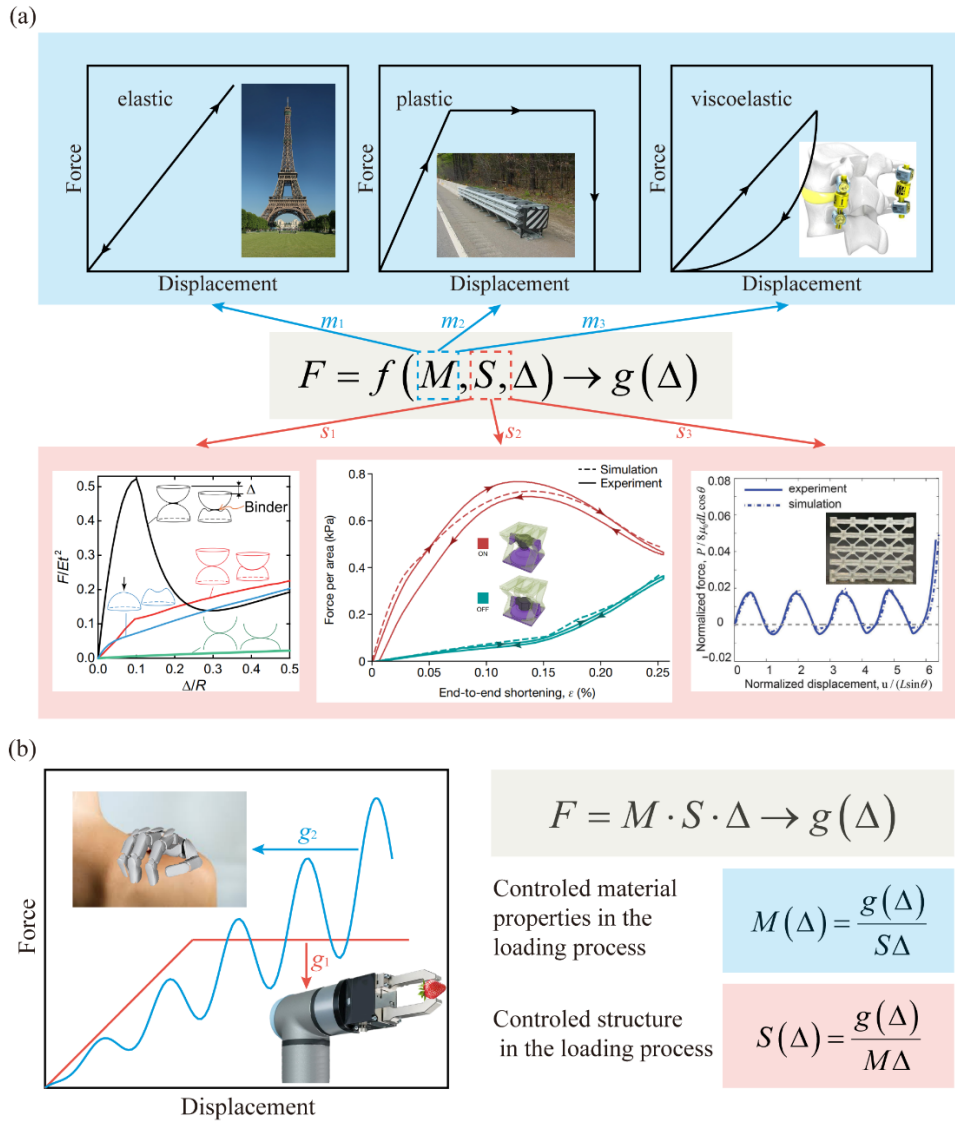

FIG.1 (a) Material property($M$) and geometry($S$) affect the force-displacement relationship of a structure. In turn, by using different materials or designing different geometries[5, 6, 17], different target force-displacement functions(*g*) can be realized. (b) The growing robotics industry has a huge demand for structures that have novel functions. The objective functions can be achieved by two approaches: tuning the material property or tuning the geometry of the structure.

## 2. Achieving objective functions by tuning the material property

In this section, a simple model of uniaxial tension, as shown in Fig.2(a), is set up to demonstrate that multiple objective force-displacement functions can be achieved by tuning the material property during the loading process. The force-displacement function of the model can be written as:

$$F = E \cdot \frac{wt}{l} \cdot \Delta \quad \rightarrow \quad g(\Delta) \tag{1}$$

Where $E$ is Young's modulus; $l$, $w$, and $t$ are the length, width, and thickness of the plate; and $g(\Delta)$ is the objective function. Normally, $E$ is considered a constant, so only linear objective functions can be achieved. But here the material property is considered tunable during the loading process, which means that $E$ is no longer a constant but a function of $\Delta$. Specifically, the material has a stiff phase with $E = 1000$MPa and a soft phase with $E = 100$MPa which are the upper and lower bounds of Young's modulus as shown in Fig.2(b). In the loading process, Young's modulus can be tuned freely between the upper and lower bounds. The force-displacement curves also have upper and lower bounds accordingly, between which, any objective functions can be achieved.

One objective function $g_1$ is proposed, as shown in Fig.2(a), which can be written as:

$$F = g(\Delta) = \begin{cases} 400 \cdot \Delta & 0 \leq \Delta < 0.5 \\ 200 & 0.5 \leq \Delta \leq 1.0 \end{cases} \tag{2}$$

To achieve this objective function, the material tuning function is calculated as:

$$E(\Delta) = \frac{g(\Delta)}{\Delta} \cdot \frac{l}{wt} = \begin{cases} 750 & 0 \leq \Delta < 0.5 \\ 375/\Delta & 0.5 \leq \Delta \leq 1.0 \end{cases} \quad (3)$$

In the loading process, if Young's modulus of the material follows the calculated tuning function, the objective force-displacement function can be achieved theoretically. To verify that, a FE model is created with the previously mentioned geometry and boundary conditions. As for the material property, a user subroutine UFIELD of ABAQUS is used to assign the material tuning function to the FE model. The force-displacement curve obtained from the simulation results is plotted in Fig.2(a), which matches well with the objective function. The stress and strain contours obtained from the simulation are shown in Fig.2(c). Obviously, with the increase of the displacement, the strain increases accordingly. However, the stress remains constant after $\Delta \geq 0.5$, which can be attributed to the decreasing of Young's modulus of the material, and therefore leads to the constant value of the force. Another two objective functions $g_2$ and $g_3$ are created and plotted in Fig.2(a) and the material tunning functions for them are calculated and plotted in Fig.2(b). FE simulations are performed and the simulation results are shown in Fig.2(a), which are also in good agreement with the objective functions.

The fundamental premise of the discussion above is that the material property can be tuned freely during the loading process. However, how to change the material property is a question that remains unanswered. Several environmental factors such as temperature, moisture, and pressure can affect the materials' properties. In particular, Young's modulus of polymer is dramatically affected by the temperature. Here, an experimental study with temperature as a tuning tool and a 3D printed polymer as a property tuning material is performed. The 3D printed tensile specimen made of VeroBlue, a photopolymer, is shown in Fig.2(d). A series of tensile tests at different temperatures are performed, and the force-displacement curves are shown in Fig.2(d). An objective force-displacement function with a flat plateau is created and shown in Fig.2(d). To achieve the objective function, a material property tuning function is calculated and therefore a temperature tuning function is calculated accordingly as shown in Fig.2(e). A heating chamber as shown in Fig.2(f) is used to control the temperature of the specimen during the loading process. In the first half part of the loading process, the temperature remains constant and then increases from 25°C to 33°C in the second half. The force-displacement curve of the test is plotted in Fig.2(d) which matches well with the objective function.

The simulation and experimental results demonstrate that objective force-displacement functions can be achieved by tuning the material property in the loading process. Moreover, the challenge of universality is also solved: multiple different objective functions can be achieved by one fabricated structure. Also, theoretically, the challenge of complexity can also be solved: any complex objective function in the design space can be achieved with a corresponding material tuning function. However, a complex objective function is hard to be achieved accurately by tuning the temperature in the loading process. Unlike a simple heating process in our test, a complex objective function may require multiple heating and cooling processes. Considering the high complexity of heat transfer, tuning the temperature of the structure accurately is practically impossible. There are two promising approaches to overcoming the challenge of complexity. One is exploring new materials whose mechanical properties can be tuned by an electric field or magnetic field quickly and accurately. Another approach is tuning the other factor that affects the relationship between force and displacement: the geometry of the structure.

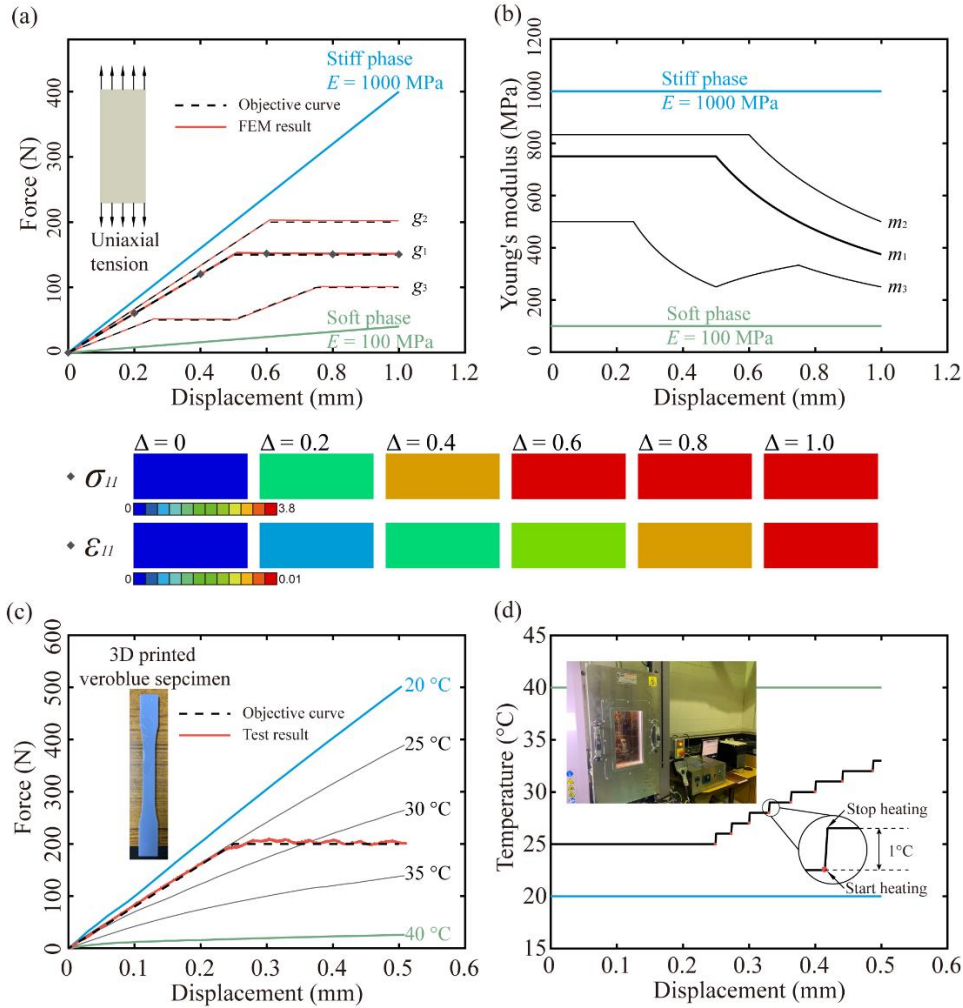

FIG.2 (a-b) A simple uniaxial tension model is assumed to have a tunable Young's modulus from 100MPa to 1000MPa. Different objective functions $g$ are proposed and corresponding material tuning functions $m$ are calculated. By applying the material tuning functions to the FE model, simulated force-displacement curves are obtained which match well with the objective functions. (c-d) The Young's modulus of a 3D printed polymer specimen is highly affected by the temperature. To achieve the objective function, a temperature tuning function is calculated are used in the experimental test to control the temperature of the specimen. The force-displacement curve from the test matches well with the objective function.

## 3. Achieving objective function by tuning the geometry of the structure

In this section, a simple model of a cantilever beam with moving support, as shown in Fig.3(a), is set up to demonstrate that multiple objective force-displacement functions can be achieved by tuning the geometry of the structure during the loading process. The force-displacement function of the model can be written as:

$$F = 3E \cdot \frac{I}{(l-x)^3} \cdot \Delta \quad \rightarrow \quad g(\Delta) \tag{4}$$

Where $E$ is Young's modulus; $I$ is the moment of inertia; $l$ is the length of the beam; $x$ is the position of the support; and $g(\Delta)$ is the objective function. Here, the cantilever beam has a moving support that can freely change its position from the left bound (green, $x = 0$) to the right bound (blue, $x = 80$) in the loading process. That is, any position function between the left and right bound in Fig.3(b) can be achieved. Therefore, the force-displacement curves in Fig.3(a) have upper and lower bounds accordingly. A complex objective function $g_1$ is proposed as shown in Fig.2(a):

$$F = g(\Delta) = \frac{\Delta}{2} \cdot (3 - \sin(\Delta \cdot \pi)) \tag{5}$$

To achieve this objective function, the support position tuning function is calculated:

$$X(\Delta) = l - \sqrt[3]{\frac{6EI}{(3-\sin(\Delta \cdot \pi))}} \qquad (6)$$

In the loading process, if the position of the moving support follows the calculated position tuning function, theoretically, the objective force-displacement function will be achieved. To verify that, a FE model is created with the previously mentioned geometry and boundary conditions and a moving support. The calculated position tuning function is applied to the moving support in the simulation. The force-displacement curve obtained from the simulation results is plotted in Fig.3(a), which matches well with the objective function. For another objective function $g_2$, position tunning function calculation and FE simulation are performed. The result is shown in Fig.3(a-b), which is also in good agreement with the objective function.

For further verification, an experimental test is performed on an aluminum cantilever beam with a moving support. The moving support is assembled on a rail and the position is controlled by a stepping motor as shown in Fig.3(c). A series of tests with the support at different positions are performed, and the force-displacement curves are shown in Fig.3(c). The position tunning function for the objective function is calculated as shown in Fig.3(d). This function is then programmed into the controller of the stepping motor and the support moves accordingly in the loading process. The force-displacement curve is obtained from the test and matches well with the objective function.

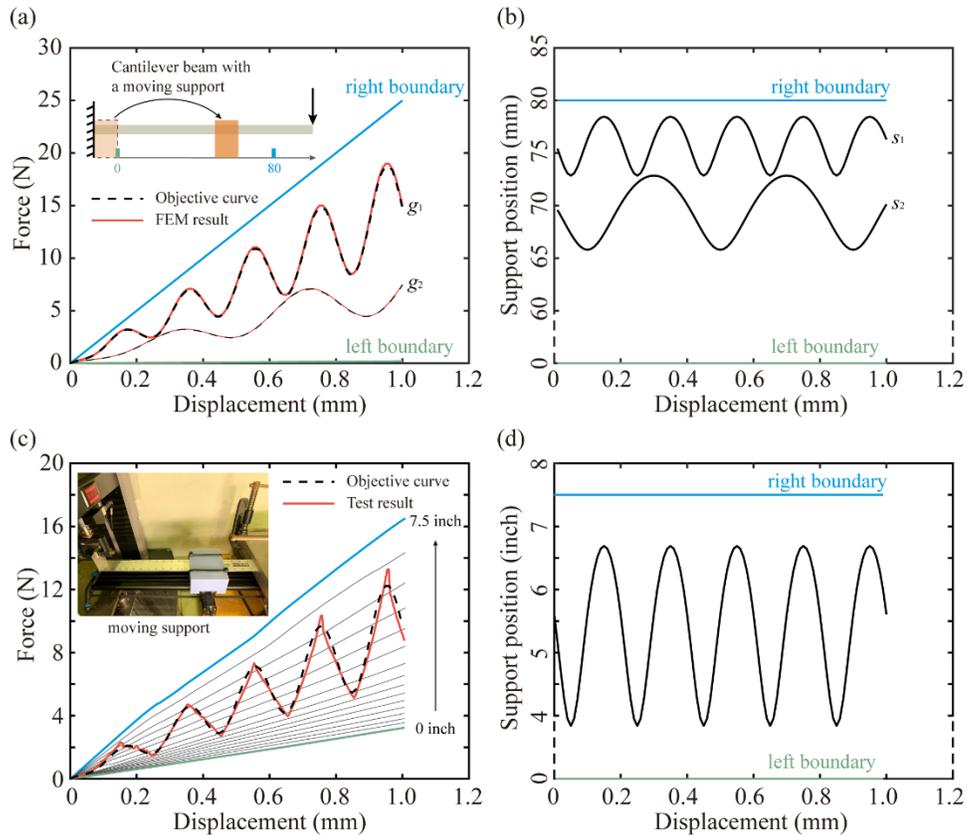

FIG.3 (a-b) A cantilever beam model with a moving support that has a left and right boundary. Different objective functions *g* are proposed and corresponding support position tuning functions *s* are calculated. By applying the support position tuning functions to the FE model, simulated force-displacement curves are obtained which match well with the objective functions. (c-d) Experimental tests are performed on an aluminum cantilever beam with a moving support. The moving support is assembled on a rail and the position is controlled by a stepping motor. To achieve the objective function, the position tuning function is calculated and used in the experimental test to control the position of the support. The force-displacement curve from the test matches well with the objective function.

The simulation and experimental results demonstrate that objective force-displacement functions can be achieved by tuning the geometry of the structure in the loading process. The two main challenges are solved at the same time: multiple different and highly complex objective functions can be achieved quickly and accurately. Based on the simple structure as shown in Fig.3(a), a truly re-programmable metamaterial is designed with four parts as shown in Fig.4(a). The stiff boundary, the deformable beam, the moving support, and the controller. Note that, the controller controls the position of the moving support during the loading process and is not necessarily a physical reality. Proper designed electric field or magnetic field that can accurately move the support can also be used as a controller.

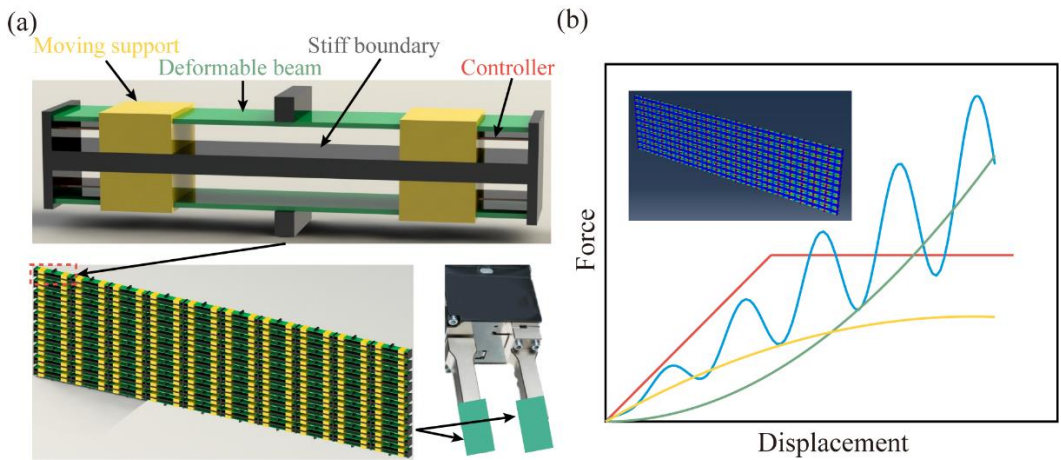

FIG.4 (a) A truly re-programmable metamaterial that can be used on robotic hands. (b) Simulation results show that multiple different and highly complex objective functions can be achieved accurately.

## 4. Conclusion

We have demonstrated two tuning approaches for a structure to achieve objective force-displacement functions. First, FE simulation and experimental test results indicated that multiple different and complex objective functions can be achieved by the material property tuning approach. However, tuning the material by temperature has several inherent drawbacks. Considering the high complexity of heat transfer, tuning the temperature of the structure quickly and accurately is practically impossible. Therefore, exploring new materials whose mechanical properties can be tuned by an electric field or magnetic field quickly and accurately is the key to practical applications.

Another tuning approach is tuning the geometry of the structure. We have presented a simple example of a cantilever beam model with a moving support. By accurately controlling the position of the support using the calculated position tuning function, both simulated and experimental results showed good agreement with the objective functions. Based on theis concept, we designed a truly re-programmable metamaterial. Multiple highly complex objective force-displacement functions were achieved. We expect that the proposed tuning approaches will facilitate the development of advanced forms of mechanical metamaterials and a new generation of robotic hands.